%% file: ms.tex
\documentclass[a4paper,cits]{PoS}
\usepackage[authoryear,square]{natbib}
\usepackage{amsmath,amssymb,graphics,graphicx}
\bibpunct{(}{)}{;}{a}{}{,}

\title{Pulsar Wind Nebulae in the SKA era}

\ShortTitle{Pulsar Wind Nebulae in the SKA era}

\author{\speaker{J. D. Gelfand}$^{ab}$, R. P. Breton$^{cf}$,
  C.-Y. Ng$^{d}$, J. W. T. Hessels$^{e}$,  B. Stappers$^{f}$,
   M. S. E. Roberts$^{a}$, A. Possenti$^{g}$ \\
\llap{$^a$}New York University Abu Dhabi, PO Box 129188, Abu Dhabi,
United Arab Emirates; E-mail: \email{joseph.gelfand@nyu.edu},
\email{mallory.roberts@nyu.edu} \\ 
\llap{$^b$}Center for Cosmology and Particle Physics, New York
University, Meyer Hall of Physics, 4 Washington Place, New York, NY
10003 \\
\llap{$^c$}School of Physics and Astronomy, University of Southampton,
  Southampton, United Kingdom; E-mail: \email{rene.breton@manchester.ac.uk} \\
\llap{$^d$}Department of Physics, The University of Hong Kong,
Pokfulam Road, Hong Kong; E-mail: \email{ncy@bohr.physics.hku.hk} \\
\llap{$^e$}ASTRON, Postbus 2, 7990 AA Dwingeloo, The Netherlands /
University of Amsterdam; E-mail: \email{hessels@astron.nl} \\
\llap{$^f$}The School of Physics and Astronomy, The University of
Manchester, Manchester, United Kingdom; E-mail:
\email{ben.stappers@manchester.ac.uk} \\
\llap{$^g$}Astronomical Observatory of Cagliari, Selargius, Italy;
E-mail: \email{possenti@oa-cagliari.inaf.it}}
  
\abstract{Neutron stars lose the bulk of their rotational energy in
  the form of a pulsar wind: an ultra-relativistic outflow of
  predominantly electrons and positrons.  This pulsar wind
  significantly impacts the environment and possible binary companion
  of the neutron star, and studying the resultant pulsar wind nebulae
  is critical for understanding the formation of neutron stars and
  millisecond pulsars, the physics of the neutron star magnetosphere,
  the acceleration of leptons up to PeV energies, and how these
  particles impact the interstellar medium.  With the SKA1 and the
  SKA2, it could be possible to study literally hundreds of PWNe in
  detail, critical for understanding the many open questions in the
  topics listed above.}

\FullConference{
Advancing Astrophysics with the Square Kilometre Array\\
June 8-13, 2014\\
Giardini Naxos, Italy}

\newcommand{\skipthis}[1]{}

\begin{document}

\input{intro.tex}
\input{nsform.tex}

\input{binaries.tex}
\input{environ.tex}
\input{requirements.tex}

\noindent {\sc Acknowledgments}: CYN is supported by a ECS grant
under HKU 709713P.  J.W.T.H. acknowledges funding from an
NWO Vidi fellowship and ERC Starting Grant "DRAGNET" (337062).

\bibliographystyle{apj_short_etal}
\bibliography{ms}


\end{document}

%% file: intro.tex
\section{Introduction}
\label{intro}
Roughly every 100 years, a neutron star is born in the Milky Way
(e.g., \citealt{faucher06}).  The rotational energy of a neutron star
powers a highly relativistic, magnetized outflow called a ``pulsar
wind'' (e.g., \citealt{goldreich69}), which creates a ``pulsar wind
nebula'' (PWN) as it expands into its surroundings.  The properties of
the PWN depend on how this wind is generated inside the neutron star's
magnetosphere, how PeV and higher energy particles are generated
inside this outflow, and the wind's interaction with its surroundings.

This is true for neutron stars young and old, isolated and in
binaries.  When the neutron star is young ($\lesssim10^4-10^5$~years
old), it is still inside the supernova remnant (SNR) created by the
progenitor explosion, and the expansion of the PWN inside the SNR
creates a ``composite SNR'' \citep{becker87}.  As described in
\S\ref{nsform}, not only do such systems allow us to study the
generation and properties of the pulsar wind for the most energetic
neutron stars, they can be used to determine the neutron star's
initial spin period and the mass and initial kinetic energy of the
supernova ejecta -- important quantities for understanding how neutron
stars are formed in these explosions.  As described in
\S\ref{binaries}, in binary systems the PWN is produced by the
interaction between the neutron star's pulsar wind and its companion.
By studying such PWNe, one can measure the properties of the pulsar
wind under very different conditions than in composite SNRs and gain
valuable insight into the last stages of the formation of millisecond
pulsars (MSPs).  Last, but not least, as described in
\S\ref{environ}, older neutron stars moving supersonically
through the interstellar medium (ISM) also produce PWNe.  Studying
these PWNe is important for understanding the magnetic field structure
of the pulsar, and how it interacts with its surroundings.

Therefore, PWNe play an important role in many areas of astrophysics --
from the explosion mechanism of core-collapse supernovae to the
physics of magnetized plasmas to the acceleration of particles in many
different physical conditions.  As described in \S\ref{requirements},
the significant improvements in collecting area and observing
capabilities promised by the SKA1 and SKA 2 has the potential
to revolutionize this field by increasing the number of well-studied
PWNe by $\sim10-100\times$, allowing for the statistical studies
needed to answer the above questions.

%% file: nsform.tex
\section{Composite Supernova Remnants}
\label{nsform}

When the pulsar wind leaves the neutron star magnetosphere, it is
thought to be a primarily equatorial outflow comprised of regions of
alternating magnetic field polarity whose energy is mainly in the
form of magnetic fields (e.g., \citealt{bogovalov99}).  The
confinement of the pulsar wind by the surrounding medium creates a
``termination shock,'' where the ``cold'' pulsar wind is converted to
a ``hot'' outflow (e.g., \citealt{kennel84}).  The PWN is then formed
by the expansion of the shocked, now particle dominated, pulsar wind
into the surrounding medium.  The significant fraction of Galactic TeV
$\gamma$-ray sources associated with PWNe ($\sim40\%$;
\citealt{tevcat}) requires that these objects produce extremely high
energy particles.  While observations support this general picture,
many basic questions remain unanswered: How is the pulsar wind
generated in the magnetosphere?  What is responsible for converting
the pulsar wind from a magnetically dominated to a particle dominated
outflow?  How are particles accelerated in these objects?

Answering these questions requires studying, in detail, the PWNe
produced by the youngest, and most energetic, neutron stars.
Understanding how particles are created in the neutron star's
magnetosphere requires measuring the total number of particles
produced by a pulsar over its lifetime.  This is possible only for
pulsars whose PWN is detected at both radio and $\gamma$-ray energies
(e.g., \citealt{dejager07, gelfand15}).  Currently, only $\sim10$
pulsars meet these criteria (e.g., \citealt{roberts04}), with the SKA
and the Cerenkov Telescope Array (CTA), this number could increase by
a factor of $\sim5-10\times$ (\S\ref{requirements}).  Ions in the
pulsar wind can explain both the high acceleration efficiency of PWNe
(e.g., \citealt{amato06}) and the detection of variable ``wisps'' near
the termination shock of several PWNe (e.g., \citealt{spitkovsky04}),
as can magnetic reconnection in the pulsar wind before (e.g.,
\citealt{kirk03}), at (e.g., \citealt{lyubarsky03}), or after the
termination shock (e.g., \citealt{porth13}).  Distinguishing between
these models requires sensitive measurements of a PWN's radio
polarization structure and spectrum (e.g., \citealt{olmi14}).
Currently, this is possible for maybe a handful of PWNe -- too few to
draw any strong conclusions.  With the SKA, it should be possible for
dozens (\S\ref{requirements}).

\begin{figure}[tb]
  \begin{center}
    \includegraphics[width=0.475\textwidth]{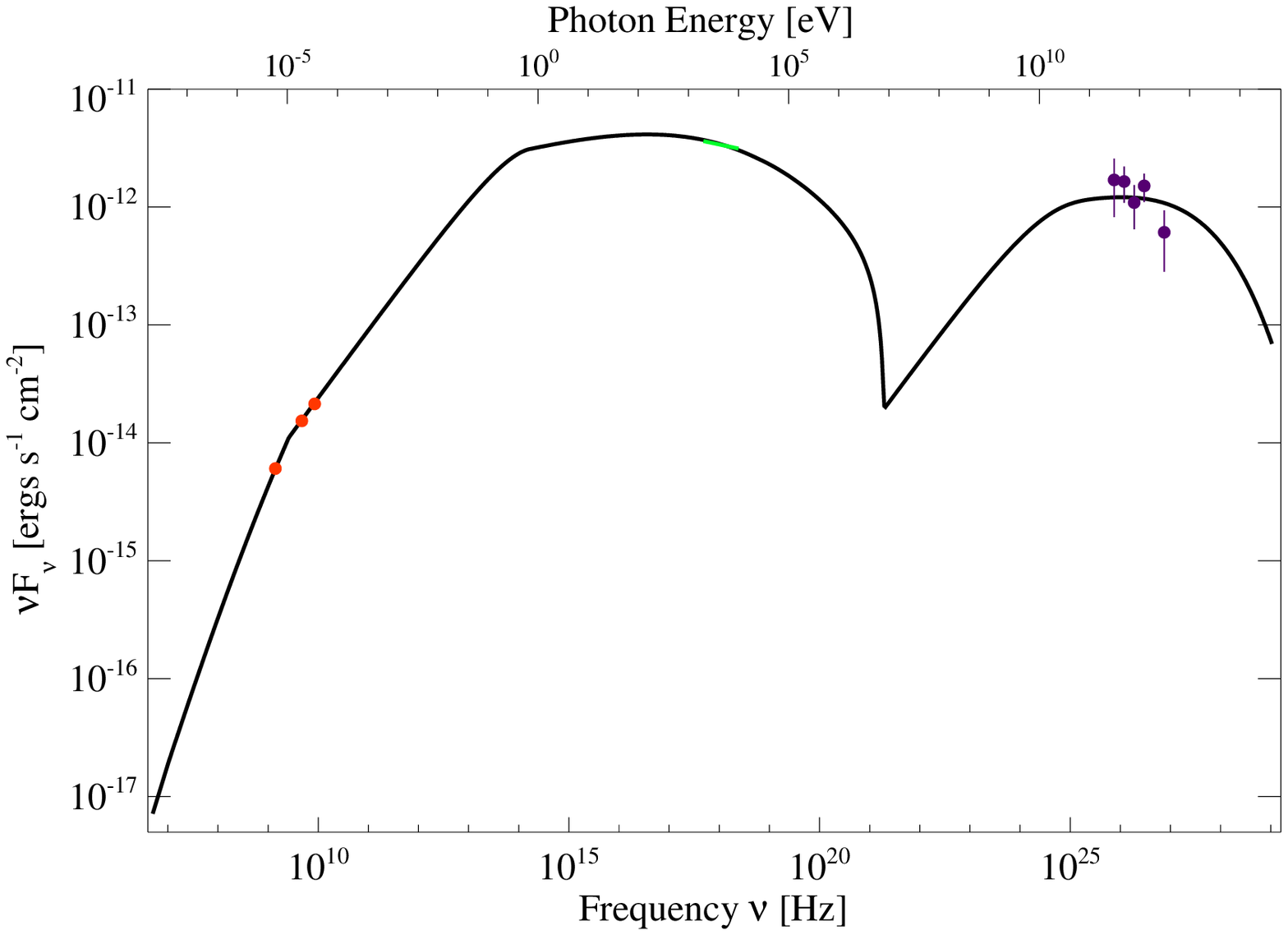}
    \includegraphics[width=0.475\textwidth]{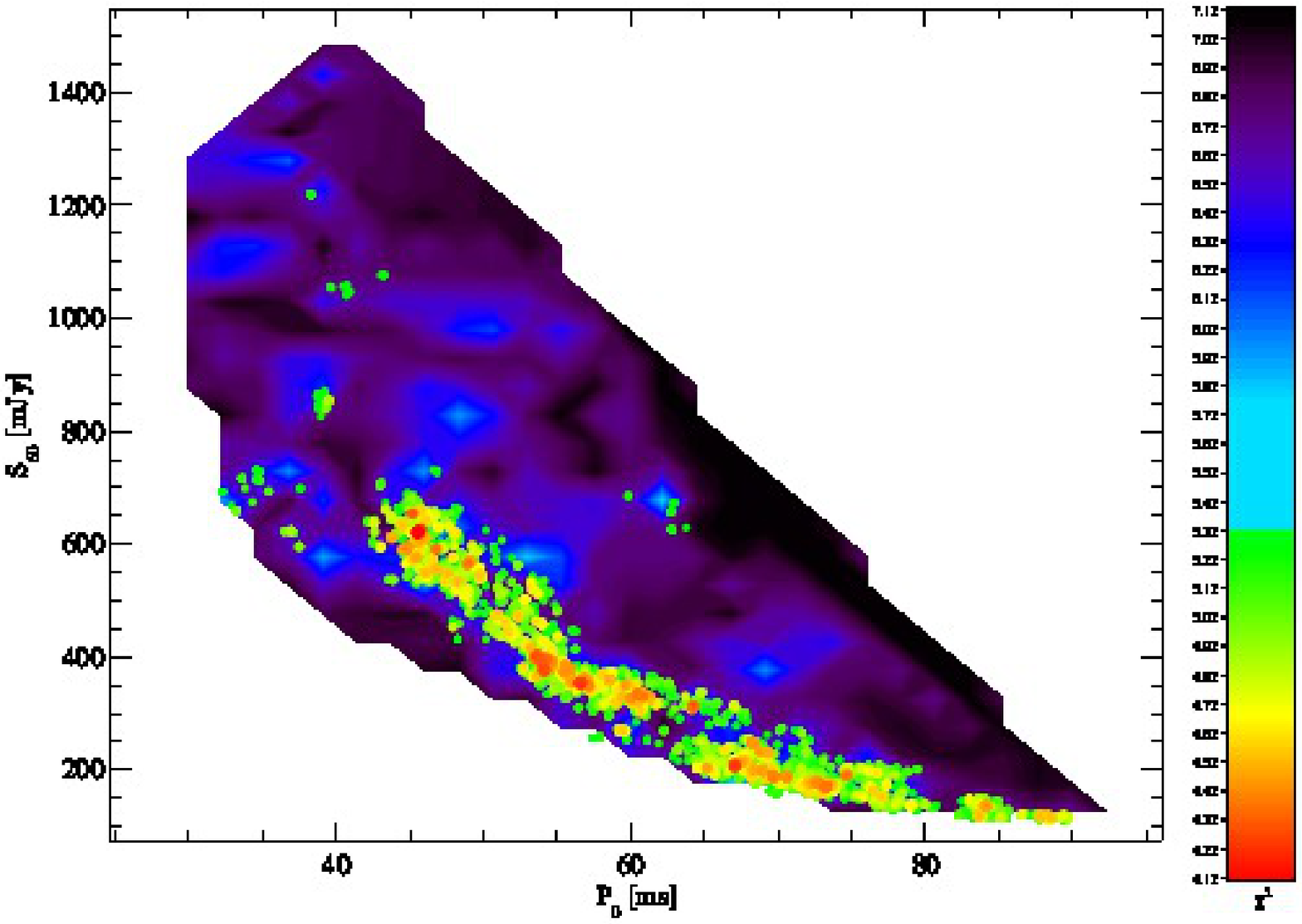}
  \end{center}
  \vspace*{-0.5cm}
  \caption{{\it Left}: Broadband SED of PWN G54.1+0.3 overlaid with
    the best fit model SED. {\it Right}: The model-predicted 60 MHz
    flux density $S_{60}$ of G54.1+0.3 as a function of the initial
    spin period $P_0$ of the central pulsar.  The color bar indicates
    the $\chi^2$ of the model fit to the observed data, with red
    (lower $\chi^2$) indicating a better fit \citep{gelfand15}.}
  \label{fig:g54model}
  \vspace*{-0.5cm}
\end{figure}

As mentioned in \S\ref{intro}, when a neutron star is young and most
energetic, it is likely inside the SNR produced by the progenitor
explosion.  In this case, the evolution of a PWN is sensitive to the
initial spin period of the central neutron star as well as the mass
and initial kinetic energy of the supernova ejecta (e.g.,
\citealt{chevalier05, gelfand09}).  This can be done using models for
the evolution of a PWN inside a SNR (Figure \ref{fig:g54model}, e.g.,
\citealt{gelfand15}), but requires measuring the current spin-down
luminosity and characteristic age of the central neutron star, the
radio, X-ray, and $\gamma$-ray spectrum of the PWN, and the radius of
the surrounding SNR.  Currently, only $\sim10$ composite SNRs meet
this criteria (e.g., \citealt{bucciantini11, torres14}), since only
$\sim50\%$ of all PWNe are associated with a pulsar (e.g.,
\citealt{roberts04}), $<50\%$ of these are associated with a SNR, and
many of the remaining PWNe detected at $\gamma$-ray energies (e.g.,
\citealt{ferrand12}) are undetected in the radio.  Even worse, a
precise estimate of the initial spin-period requires measuring the
braking index of the central pulsar (e.g., \citealt{gelfand14}),
currently accomplished for $<5$ sources.  As described in
\S\ref{requirements}, the SKA could increase this number by
$\sim5-10\times$, enabling one to directly test different models for
the formation of neutron stars in core-collapse supernovae (e.g.,
\citealt{watts02, blondin07}).

Lastly, as mentioned above, PWNe dominate the luminous TeV
$\gamma$-ray population of the Milky Way \citep{carrigan13}, with TeV
PWNe typically associated with pulsars whose characteristic ages are
$t_{\rm ch} \lesssim10^5$~years \citep{tevcat}. Therefore, the
majority of these pulsars, and PWNe, will be inside the SNR of their
progenitor explosion even if this SNR is not detected.  With the SKA1
and SKA2, we will determine if this is true for the considerable
number of currently unidentified Milky Way TeV sources, by discovering
a coincident ``young'' radio pulsar, diffuse, flat spectrum radio
emission characteristic of a PWN, and/or a surrounding steep-spectrum
shell suggestive of a SNR.  Additionally, the SKA1 and SKA2 will
discover PWNe around energetic pulsars whose radio beams do {\it not}
point towards the Earth.  For all these objects, the flux densities
measured by SKA1-LOW will allow us to estimate their initial periods
even if pulsed emission is not detected.

%% file: binaries.tex
\section{Neutron Star Binaries}
\label{binaries}

\begin{figure}[tb]
  \vspace*{-4cm}
  \begin{center}
    \includegraphics[width=0.5\textwidth]{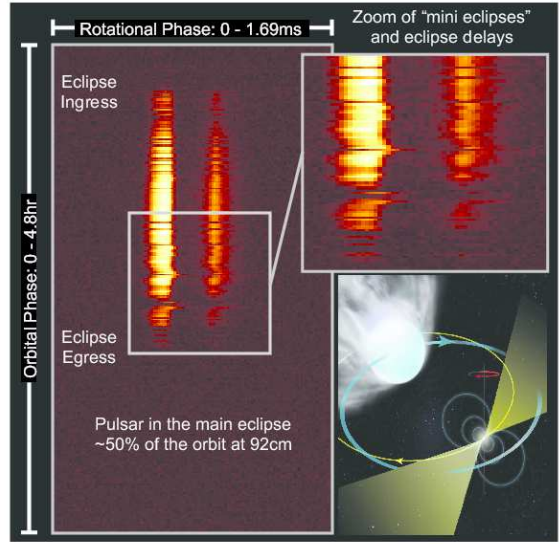}  
  \end{center}
  \vspace*{-1.0cm}
  \caption{Eclipses in the pulsed radio emission from PSR~J1023+0038
    (Hessels, priv. comm.) as well as an artist's impression of the
    cloud of plasma resulting from the ablation of its low-mass
    companion by its pulsar wind (Bill Saxton, NRAO).}
  \label{fig:eclipses}
  \vspace*{-0.5cm}
\end{figure}

If a neutron star is in a binary system, the highly relativistic
``pulsar wind'' powered by its rotational energy will interact with
its companion.  The interaction between the neutron star's pulsar wind
and its stellar companion leads to an intrabinary shock which possibly
accelerates particles to high energies (e.g., \citealt{bogdanov11}),
filling the system with plasma.  Evidence for plasma production in
pulsar binaries has been observed from high-mass X-ray binaries (e.g.,
\citealt{moldon14}) to ``black widow'' and ``redback'' systems (e.g.,
\citealt{roberts14}), where the pulsar wind actually ablates material
from the companion star (Figure \ref{fig:eclipses}, e.g.,
\citealt{fruchter88}).  Studying the intrabinary plasma probes the
pulsar wind much closer to the neutron star magnetosphere
\citep{petri11} than the systems described in \S\ref{nsform}, where
extremely different conditions may prevail.  In fact, analysis of one
binary pulsar system suggests the wind is strongly magnetized in this
regime \citep{bogdanov11}, while the study of the PWNe described in
\S\ref{nsform} require a weakly magnetized wind further from the
neutron star.  If correct, this places strong constraints on models
for magnetic reconnection in the pulsar wind (\S\ref{nsform}), the
currently leading theory for particle acceleration in these systems
(e.g. \citealt{kirk03}).

Typically, such studies require detecting orbitally-modulated
non-thermal X-ray emission from the binary (e.g.,
\citealt{bogdanov11}), not achievable with a radio telescope like the
SKA.  However, as described in \S\ref{requirements}, the collecting
area of the SKA2 could increase by $>10\times$ the number of binary
pulsars -- providing much needed targets for these X-ray studies.
Additionally, inhomogeneities and/or free-free absorption in the cloud
of plasma generated at the intrabinary shock is believed to be
responsible for the ``eclipses'' in the pulsed radio emission
(Fig. \ref{fig:eclipses}) observed from $\sim50$ binaries pulsars
\citep{roberts13,friere13}.  By simultaneously measuring the pulsed
and unpulsed continuum radio flux of these systems, it is possible to
determine the density, geometry, and filling factor of this plasma as
well as the structure of its magnetic field -- especially if there are
simultaneous or contemporaneous $\gamma$-ray observations.  With the
SKA1 and SKA2, we expect to make such measurements for $\sim300$
eclipsing binaries (\S\ref{requirements}), allowing one to determine
how the properties of the plasma produced in this interaction depend
on the characteristics of the neutron star and binary system (e.g.,
neutron star spin-down luminosity, binary separation, companion mass
and radius).

Lastly, the recent discovery of several millisecond pulsars (MSPs) /
low mass X-ray binaries (LMXBs) transitioning between an
``accretion''-dominated phase where no radio pulsations are detected
and a MSP phase which shows no evidence for accretion (e.g.,
\citealt{archibald09, bassa14, papitto13, stappers14}) promises
important insight into the formation of MSPs as well as understanding
accretion onto magnetized neutron stars.  Since studies of
intermittent pulsars strongly suggest that the emission of radio
pulses is connected to their loss of rotational energy (e.g.,
\citealt{young13,li14}), the physical mechanism responsible for
generating a pulsar wind likely plays an important role in this
transition.  Only by studying the radio and higher energy (X-ray,
$\gamma$-ray) properties of such systems will it be possible to
understand the interplay between accretion and a neutron star's
magnetosphere.  As described in \S\ref{requirements}, the potential
multi-beam and high-frequency capabilities of the SKA are critical for
understanding the physics of these transitions which may comprise the
final stage in the formation of a MSP (though some systems might be
stuck transitioning back-and-forth to a LMXB state).

%% file: environ.tex
\section{Isolated Pulsars in the Interstellar Medium}
\label{environ}

The pulsar winds of isolated, older neutron stars will also
significantly impact their surrounding.  In this case, the pulsar wind
is confined by the ram pressure created by the neutron star's
supersonic motion.  The morphology (e.g., \citealt{vigelius07}) and
the magnetic field structure (e.g., \citealt{bucciantini05a}) of such
PWNe are sensitive to the density structure of the ISM, the speed and
direction of the neutron star's spatial velocity, and the geometry of
its pulsar wind.  Additionally, X-ray observations suggest such PWNe
efficiently inject high-energy particles into the surrounding ISM,
often in directions misaligned with the neutron star's proper motion
(e.g., \cite{bandiera08, deluca13, marelli13, pavan14}).

\begin{figure}[tb]
  \vspace*{-0.5cm}
  \begin{center}
    \includegraphics[width=0.75\textwidth]{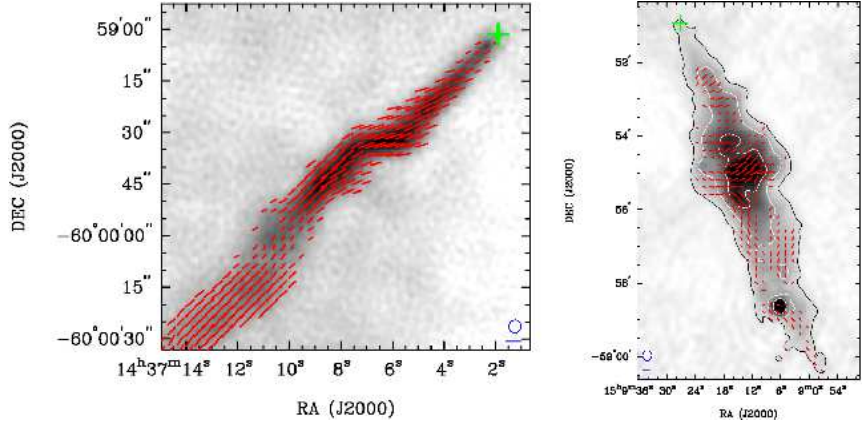}
  \end{center}
  \vspace*{-0.5cm}
  \caption{The radio nebulae G315.9$-$0.0 ({\it left}; \citealt{ng12})
    and G319.9$-$0.7 ({\it right}; \citealt{ng10}.  The location of
    the associated pulsars are indicated by the green crosses, while
    the polarization $\vec{B}$-vector is shown by the red lines.}
  \label{fig:pol}
  \vspace*{-0.5cm}
\end{figure}

Understanding the structure of these pulsar winds, and how high energy
particles escape the PWN, requires mapping the polarized radio
emission for a range of neutron star and ISM properties.  Currently,
this is possible for $\lesssim5-10$ systems, but even this small
sample shows significant diversity in their magnetic field structure
-- for example, the ``Frying Pan'' (G315.9$-$0.0) PWN has a magnetic
field aligned with the pulsar's proper motion \citep{ng12}, while the
magnetic field of G319.9$-$0.7 has a helical structure \cite{ng10}
(Figure \ref{fig:pol}) -- suggesting a dependence on the flow
conditions, pulsar speed, and/or misalignment between the pulsar
rotation axis and proper motion.  Understanding the relationship
between these different physical parameters requires measuring the
magnetic field structure of many more PWN which, as described in
\S\ref{requirements}, will become possible with the SKA.

%% file: requirements.tex
\section{Requirements for SKA1 and SKA2}
\label{requirements}
Thanks to the considerable improvement in sensitivity promised by the
SKA1 and the SKA2, these instruments should be capable of measuring
the radio morphology, spectrum, and polarization properties of
$\sim100$ PWNe -- a significant improvement over the $\sim10$ for
which such measurements exist, and increase the number of eclipsing
binary pulsars from $\sim50$ to $\sim300$.  Such sample sizes are
needed to determine how a neutron star's spin-down luminosity, age,
and environment affects the evolution of its PWN.  Additionally, the
SKA1 and especially SKA2 will be able to measure properties of a PWN
never before possible.  These studies depend less on the collecting
area of the SKA1, and therefore are more immune to a 50\% reduction of
its capabilities, but on the configuration and capabilities of its
telescopes as described below:

\paragraph{Discover pulsars:} Currently, $\sim40-50\%$ of PWN
candidates are unassociated with a pulsar in any waveband
\citep{roberts04}, problematic for the analyses discussed in
\S\ref{nsform} and \S\ref{environ}.  While beaming explains some of
the missing pulsars, for many PWNe it likely results from the
limitations of current observatories.  The large collecting area of
the SKA1-LOW should increase the number of pulsars detected in nearby
PWN, but the high dispersion measure (DM) and scattering timescales
expected for pulsars on the far side of the Milky Way require that the
SKA1-MID also be able to detect pulsars.

\paragraph{Monitor a PWN's pulsed and unpulsed radio emission:}
Monitoring the timing properties of young, isolated pulsars is needed
to measure their braking indices which, as described in \S\ref{nsform},
is critical for estimating their initial spin period.  For eclipsing
binary pulsars, monitoring the pulsed emission allows one to detect
changes in the orbital phase and length of radio eclipses changes, and
monitoring the pulsed emission of MSP/LMXB is critical for determining
if the pulsed radio emission disappears before or after the resumption
of accretion onto the neutron star (\S\ref{binaries}).  Both the
SKA1-LOW and SKA1-MID require these capabilities since the possible
presence of dense plasma in these systems will result in frequency
dependent behavior.  Additionally, observations of young PWNe every
few months are required to measure the variability of ``wisps'' near
the termination shock, critical for determining if ions are present in
the pulsar wind (\S\ref{nsform}).  Since these wisps are only a few
arcseconds in size (e.g., \citealt{bietenholz04}), this is best done
with the higher frequencies of SKA1-MID.

\paragraph{Conduct pulsar gating and VLBI observations:} While all
pulsars are believed to generate a PWN, radio PWNe are detected around
$<10\%$ of young, energetic pulsars.  One possibility is that the
radio PWN is masked in continuum observations by the pulsar's pulsed
emission.  With pulsar gating, it is possible to image the region when
the pulsar is ``off'', making it possible to detect faint radio PWNe
\citep{gaensler98}.  This is particularly important for SKA1-LOW, due
to its lower angular resolution and increased brightness of the pulsar
relative to the PWN.  Pulsar gating of eclipsing binary pulsars allows
simultaneous measurements of its pulsed and unpulsed flux, critical
for determining the eclipsing mechanism (\S\ref{binaries}), and
significantly improves measurements of a pulsar's parallax and proper
motion in VLBI observations -- important for measuring the initial
spin periods of young pulsars in composite SNRs (e.g.,
\citealt{gelfand15}) and interpreting the morphology of ``bow shock''
PWNe (\S\ref{environ}).

\paragraph{Observe PWNe across a broad range of frequencies:} The
broadband radio spectrum of a PWN inside a SNR is expected to have
numerous features (e.g., \citealt{gelfand09}).  The minimum particle
energy injected at the termination shock will result in a ``break'' at
low ($\nu \lesssim 1~{\rm GHz}$) frequencies, and measuring its flux
density below this break is critical for estimating the initial spin
period of the central neutron star (e.g., \citealt{gelfand15}; Figure
\ref{fig:g54model}).  Ions in the pulsar wind and/or magnetic
reconnection downstream of the termination shock (\S\ref{nsform}) is
expected to lead to broad ``bumps'' in the spectrum at higher
frequencies ($\nu \gtrsim 10~{\rm GHz}$; \citealt{olmi14}).
Continuous frequency coverage between the different SKA1 and SKA2
observing bands would improve measurements of curvature in the PWN's
radio spectrum.

\paragraph{Detect emission on large angular scales:} The non-detection
of many TeV PWNe at radio wavelengths (e.g., \citealt{ferrand12}) and
SNRs around young pulsars / PWNe (e.g., \citealt{roberts04}) likely
results from the large angular size ($\sim30^\prime -
\lesssim1^{\circ}$) and low radio surface brightness of these
objects.  Current radio interferometers do not have the short
baselines needed to detect emission on these angular scales, and
single-dish telescopes do not have the sensitivity needed to detect
this emission over the Galactic background.  While SKA1-LOW and
SKA1-MID will have the needed sensitivity, a dense core is required to
detect emission on the needed angular scales.
instantaneous SKA1-LOW single beam

\paragraph{Measure the polarization properties of PWNe:} Last, but not
least, mapping the magnetic field structure of young
(\S\ref{nsform}) and old (\S\ref{environ}) PWNe requires spatially resolved
measurements of their polarized intensity and direction at multiple
frequencies to both correct for foreground Faraday rotation and detect
changes in its rotation measure (e.g., \citealt{ng10}).  Furthermore,
measuring changes in the DM during the ``eclipse'' of a binary pulsar
(e.g., \citealt{archibald13}) are important for measuring the density
and magnetic field structure of the intervening plasma
(\S\ref{binaries}).  These observations require high polarization
purity of both SKA1-LOW and SKA1-MID, and sensitivity to polarized
emission over a small bandwidth or at higher ($\nu > 1~{\rm GHz}$)
frequencies to avoid bandwidth depolarization, especially for
SKA1-MID.  Furthermore, since a PWN's radio emission can be highly
linearly polarized, an all-sky polarization survey with the SKA1-SUR
could identify new PWNe, particularly around older, isolated pulsars.

%% file: ms.bbl
\begin{thebibliography}{48}
\expandafter\ifx\csname natexlab\endcsname\relax\def\natexlab#1{#1}\fi

\bibitem[{{Amato} \& {Arons}(2006)}]{amato06}
{Amato}, E. \& {Arons}, J. 2006, \apj, 653, 325

\bibitem[{{Archibald} {et~al.}(2013){Archibald}, {Kaspi}, {Hessels},
  {Stappers}, {Janssen}, \& {Lyne}}]{archibald13}
{Archibald}, A.~M. \etal . 2013, ArXiv e-prints

\bibitem[{{Archibald} {et~al.}(2009){Archibald}, {Stairs}, {Ransom}, {Kaspi},
  {Kondratiev}, {Lorimer}, {McLaughlin}, {Boyles}, {Hessels}, {Lynch}, {van
  Leeuwen}, {Roberts}, {Jenet}, {Champion}, {Rosen}, {Barlow}, {Dunlap}, \&
  {Remillard}}]{archibald09}
---. 2009, Science, 324, 1411

\bibitem[{{Bandiera}(2008)}]{bandiera08}
{Bandiera}, R. 2008, \aap, 490, L3

\bibitem[{{Bassa} {et~al.}(2014){Bassa}, {Patruno}, {Hessels}, {Keane},
  {Monard}, {Mahony}, {Bogdanov}, {Corbel}, {Edwards}, {Archibald}, {Janssen},
  {Stappers}, \& {Tendulkar}}]{bassa14}
{Bassa}, C.~G. \etal . 2014, \mnras, 441, 1825

\bibitem[{{Bietenholz} {et~al.}(2004){Bietenholz}, {Hester}, {Frail}, \&
  {Bartel}}]{bietenholz04}
{Bietenholz}, M.~F. \etal . 2004, \apj, 615, 794

\bibitem[{{Blondin} \& {Mezzacappa}(2007)}]{blondin07}
{Blondin}, J.~M. \& {Mezzacappa}, A. 2007, \nat, 445, 58

\bibitem[{{Bogdanov} {et~al.}(2011){Bogdanov}, {Archibald}, {Hessels}, {Kaspi},
  {Lorimer}, {McLaughlin}, {Ransom}, \& {Stairs}}]{bogdanov11}
{Bogdanov}, S. \etal . 2011, \apj, 742, 97

\bibitem[{{Bogovalov}(1999)}]{bogovalov99}
{Bogovalov}, S.~V. 1999, \aap, 349, 1017

\bibitem[{{Bucciantini} {et~al.}(2005){Bucciantini}, {Amato}, \& {Del
  Zanna}}]{bucciantini05a}
{Bucciantini}, N. \etal . 2005, \aap, 434, 189

\bibitem[{{Bucciantini} {et~al.}(2011){Bucciantini}, {Arons}, \&
  {Amato}}]{bucciantini11}
---. 2011, \mnras, 410, 381

\bibitem[{{Carrigan} {et~al.}(2013){Carrigan}, {Brun}, {Chaves}, {Deil},
  {Donath}, {Gast}, {Marandon}, {Renaud}, \& {for the
  H.~E.~S.~S.~collaboration}}]{carrigan13}
{Carrigan}, S. \etal . 2013, ArXiv e-prints

\bibitem[{{Chevalier}(2005)}]{chevalier05}
{Chevalier}, R.~A. 2005, \apj, 619, 839

\bibitem[{{de Jager}(2007)}]{dejager07}
{de Jager}, O.~C. 2007, \apj, 658, 1177

\bibitem[{{De Luca} {et~al.}(2013){De Luca}, {Mignani}, {Marelli}, {Salvetti},
  {Sartore}, {Belfiore}, {Saz Parkinson}, {Caraveo}, \& {Bignami}}]{deluca13}
{De Luca}, A. \etal . 2013, \apjl, 765, L19

\bibitem[{{Faucher-Gigu{\`e}re} \& {Kaspi}(2006)}]{faucher06}
{Faucher-Gigu{\`e}re}, C.-A. \& {Kaspi}, V.~M. 2006, \apj, 643, 332

\bibitem[{{Ferrand} \& {Safi-Harb}(2012)}]{ferrand12}
{Ferrand}, G. \& {Safi-Harb}, S. 2012, Advances in Space Research, 49, 1313

\bibitem[{{Freire}(2013)}]{friere13}
{Freire}, P. 2013, http://www.naic.edu/~pfreire/GCpsr.html

\bibitem[{{Fruchter} {et~al.}(1988){Fruchter}, {Stinebring}, \&
  {Taylor}}]{fruchter88}
{Fruchter}, A.~S. \etal . 1988, \nat, 333, 237

\bibitem[{{Gaensler} {et~al.}(1998){Gaensler}, {Stappers}, {Frail}, \&
  {Johnston}}]{gaensler98}
{Gaensler}, B.~M. \etal . 1998, \memsai, 69, 813

\bibitem[{{Gelfand} {et~al.}(2014){Gelfand}, {Slane}, \& {Temim}}]{gelfand14}
{Gelfand}, J.~D. \etal . 2014, Astronomische Nachrichten, 335, 318

\bibitem[{{Gelfand} {et~al.}(submitted){Gelfand}, {Slane}, \&
  {Temim}}]{gelfand15}
---. submitted, \apj

\bibitem[{{Gelfand} {et~al.}(2009){Gelfand}, {Slane}, \& {Zhang}}]{gelfand09}
---. 2009, \apj, 703, 2051

\bibitem[{{Goldreich} \& {Julian}(1969)}]{goldreich69}
{Goldreich}, P. \& {Julian}, W.~H. 1969, \apj, 157, 869

\bibitem[{{Helfand} \& {Becker}(1987)}]{becker87}
{Helfand}, D.~J. \& {Becker}, R.~H. 1987, \apj, 314, 203

\bibitem[{{Kennel} \& {Coroniti}(1984)}]{kennel84}
{Kennel}, C.~F. \& {Coroniti}, F.~V. 1984, \apj, 283, 694

\bibitem[{{Kirk} \& {Skj{\ae}raasen}(2003)}]{kirk03}
{Kirk}, J.~G. \& {Skj{\ae}raasen}, O. 2003, \apj, 591, 366

\bibitem[{{Li} {et~al.}(2014){Li}, {Tong}, {Yan}, {Yuan}, {Xu}, \&
  {Wang}}]{li14}
{Li}, L. \etal . 2014, \apj, 788, 16

\bibitem[{{Lyubarsky}(2003)}]{lyubarsky03}
{Lyubarsky}, Y.~E. 2003, \mnras, 345, 153

\bibitem[{{Marelli} {et~al.}(2013){Marelli}, {De Luca}, {Salvetti}, {Sartore},
  {Sartori}, {Caraveo}, {Pizzolato}, {Saz Parkinson}, \&
  {Belfiore}}]{marelli13}
{Marelli}, M. \etal . 2013, \apj, 765, 36

\bibitem[{{Mold{\'o}n} {et~al.}(2014){Mold{\'o}n}, {Rib{\'o}}, {Paredes}, \&
  {Johnston}}]{moldon14}
{Mold{\'o}n}, J. \etal . 2014, International Journal of Modern Physics
  Conference Series, 28, 60173

\bibitem[{{Ng} {et~al.}(2012){Ng}, {Bucciantini}, {Gaensler}, {Camilo},
  {Chatterjee}, \& {Bouchard}}]{ng12}
{Ng}, C.-Y. \etal . 2012, \apj, 746, 105

\bibitem[{{Ng} {et~al.}(2010){Ng}, {Gaensler}, {Chatterjee}, \&
  {Johnston}}]{ng10}
---. 2010, \apj, 712, 596

\bibitem[{{Olmi} {et~al.}(2014){Olmi}, {Del Zanna}, {Amato}, {Bandiera}, \&
  {Bucciantini}}]{olmi14}
{Olmi}, B. \etal . 2014, \mnras, 438, 1518

\bibitem[{{Papitto} {et~al.}(2013){Papitto}, {Ferrigno}, {Bozzo}, {Rea},
  {Pavan}, {Burderi}, {Burgay}, {Campana}, {di Salvo}, {Falanga},
  {Filipovi{\'c}}, {Freire}, {Hessels}, {Possenti}, {Ransom}, {Riggio},
  {Romano}, {Sarkissian}, {Stairs}, {Stella}, {Torres}, {Wieringa}, \&
  {Wong}}]{papitto13}
{Papitto}, A. \etal . 2013, \nat, 501, 517

\bibitem[{{Pavan} {et~al.}(2014){Pavan}, {Bordas}, {P{\"u}hlhofer},
  {Filipovi{\'c}}, {De Horta}, {O'Brien}, {Balbo}, {Walter}, {Bozzo},
  {Ferrigno}, {Crawford}, \& {Stella}}]{pavan14}
{Pavan}, L. \etal . 2014, \aap, 562, A122

\bibitem[{{P{\'e}tri} \& {Dubus}(2011)}]{petri11}
{P{\'e}tri}, J. \& {Dubus}, G. 2011, \mnras, 417, 532

\bibitem[{{Porth} {et~al.}(2013){Porth}, {Komissarov}, \& {Keppens}}]{porth13}
{Porth}, O. \etal . 2013, \mnras, 431, L48

\bibitem[{{Roberts}(2004)}]{roberts04}
{Roberts}, M.~S.~E. 2004, http://www.physics.mcgill.ca/~pulsar/pwncat.html

\bibitem[{{Roberts}(2013)}]{roberts13}
{Roberts}, M.~S.~E. 2013, in IAU Symposium, Vol. 291, IAU Symposium, ed.
  J.~{van Leeuwen}, 127--132

\bibitem[{{Roberts} {et~al.}(2014){Roberts}, {Mclaughlin}, {Gentile}, {Aliu},
  {Hessels}, {Ransom}, \& {Ray}}]{roberts14}
{Roberts}, M.~S.~E. \etal . 2014, Astronomische Nachrichten, 335, 313

\bibitem[{{Spitkovsky} \& {Arons}(2004)}]{spitkovsky04}
{Spitkovsky}, A. \& {Arons}, J. 2004, \apj, 603, 669

\bibitem[{{Stappers} {et~al.}(2014){Stappers}, {Archibald}, {Hessels}, {Bassa},
  {Bogdanov}, {Janssen}, {Kaspi}, {Lyne}, {Patruno}, {Tendulkar}, {Hill}, \&
  {Glanzman}}]{stappers14}
{Stappers}, B.~W. \etal . 2014, \apj, 790, 39

\bibitem[{{Torres} {et~al.}(2014){Torres}, {Cillis}, {Mart{\'{\i}}n}, \& {de
  O{\~n}a Wilhelmi}}]{torres14}
{Torres}, D.~F. \etal . 2014, Journal of High Energy Astrophysics, 1, 31

\bibitem[{{Vigelius} {et~al.}(2007){Vigelius}, {Melatos}, {Chatterjee},
  {Gaensler}, \& {Ghavamian}}]{vigelius07}
{Vigelius}, M. \etal . 2007, \mnras, 374, 793

\bibitem[{{Wakely} \& {Horan}(2008)}]{tevcat}
{Wakely}, S.~P. \& {Horan}, D. 2008, International Cosmic Ray Conference, 3,
  1341

\bibitem[{{Watts} \& {Andersson}(2002)}]{watts02}
{Watts}, A.~L. \& {Andersson}, N. 2002, \mnras, 333, 943

\bibitem[{{Young} {et~al.}(2013){Young}, {Stappers}, {Lyne}, {Weltevrede},
  {Kramer}, \& {Cognard}}]{young13}
{Young}, N.~J. \etal . 2013, \mnras, 429, 2569

\end{thebibliography}
